\documentstyle[12pt,twoside,psfig]{article}

\def\be{\begin{equation}}
\def\ee{\end{equation}}
\def\bea{\begin{eqnarray}}
\def\ena{\end{eqnarray}}

\def\a{\alpha}
\def\b{\beta}

\def\d{\delta}
\def\e{\epsilon}           % Also, \varepsilon
\def\h{\eta}

\def\j{\psi}
\def\p{\pi}                % Also, \varpi
\def\q{\theta}                    %     \vartheta
\def\s{\sigma}                    %     \varsigma
\def\t{\tau}

\def\z{\zeta}
\def\D{\Delta}

\def\G{\Gamma}

\def\O{\Omega}

\def\del{\partial}

\catcode`@=11

\def\un#1{\relax\ifmmode\@@underline#1\else
$\@@underline{\hbox{#1}}$\relax\fi}

{\catcode`\'=\active \gdef'{{}^\bgroup\prim@s}}

\def\magstep#1{\ifcase#1 \@m\or 1200\or 1440\or 1728\or 2074\or 2488\or
       2986\fi\relax}   % to include magstep6

\catcode`@=12

\def\bop#1{\setbox0=\hbox{$#1M$}\mkern1.5mu
	\vbox{\hrule height0pt depth.04\ht0
	\hbox{\vrule width.04\ht0 height.9\ht0 \kern.9\ht0
	\vrule width.04\ht0}\hrule height.04\ht0}\mkern1.5mu}
\def\pa{\partial}                              % curly d
\def\leftrightarrowfill{$\mathsurround=0pt \mathord\leftarrow \mkern-6mu
       \cleaders\hbox{$\mkern-2mu \mathord- \mkern-2mu$}\hfill
       \mkern-6mu \mathord\rightarrow$}
\def\dvec#1{\vbox{\ialign{##\crcr
       \leftrightarrowfill\crcr\noalign{\kern-1pt\nointerlineskip}
       $\hfil\displaystyle{#1}\hfil$\crcr}}}          % <--> accent
\def\hook#1{{\vrule height#1pt width0.4pt depth0pt}}
\def\leftrighthookfill#1{$\mathsurround=0pt \mathord\hook#1
       \hrulefill\mathord\hook#1$}
\def\underhook#1{\vtop{\ialign{##\crcr                 % |_| under
       $\hfil\displaystyle{#1}\hfil$\crcr
       \noalign{\kern-1pt\nointerlineskip\vskip2pt}
       \leftrighthookfill5\crcr}}}
\def\smallunderhook#1{\vtop{\ialign{##\crcr      % " for su'scripts
       $\hfil\scriptstyle{#1}\hfil$\crcr
       \noalign{\kern-1pt\nointerlineskip\vskip2pt}
       \leftrighthookfill3\crcr}}}
\def\sfrac#1#2{{\vphantom1\smash{\lower.5ex\hbox{\small$#1$}}\over
       \vphantom1\smash{\raise.4ex\hbox{\small$#2$}}}} % alt. fraction
\def\bfrac#1#2{{\vphantom1\smash{\lower.5ex\hbox{$#1$}}\over
       \vphantom1\smash{\raise.3ex\hbox{$#2$}}}}      % "
\def\afrac#1#2{{\vphantom1\smash{\lower.5ex\hbox{$#1$}}\over#2}}  %"
\def\on#1#2{{\buildrel{\mkern2.5mu#1\mkern-2.5mu}\over{#2}}}%acc.over
\def\ddt#1{\on{\hbox{\LARGE .\kern-2pt.}}#1}             % double dot
\def\tdt#1{\on{\hbox{\LARGE .\kern-2pt.\kern-2pt.}}#1}   % triple dot

\def\boxes#1{
       \newcount\num
       \num=1
       \newdimen\downsy
       \downsy=-1.5ex
       \mskip-2.8mu
       \bo
       \loop
       \ifnum\num<#1
       \llap{\raise\num\downsy\hbox{$\bo$}}
       \advance\num by1
       \repeat}
\def\boxup#1#2{\newcount\numup
       \numup=#1
       \advance\numup by-1
       \newdimen\upsy
       \upsy=.75ex
       \mskip2.8mu
       \raise\numup\upsy\hbox{$#2$}}

\newskip\humongous \humongous=0pt plus 1000pt minus 1000pt

\newif\ifdtup

\parskip=\medskipamount  % space between paragraphs (TeX)
\def\baselinestretch{1.2}% magnification for line spacing (LaTeX)
\thicklines              % thick straight lines for pictures (LaTeX)
\def\border{
 }
\def\headpic{
 }% SB paper head
\def\title#1#2#3#4{\begin{document}
       \border
       \headpic
       {\hbox to\hsize{#4 \hfill ITP-SB-#3}}\par
       \begin{center}\vskip.8in minus.1in
       {\Large\bf #1}\\[.5in minus.2in]{#2}
       \vskip1.4in minus1.2in {\bf ABSTRACT}\\[.1in]\end{center}
       \begin{quotation}\par}
\def\author#1#2{#1\\[.1in]{\it #2}\\[.1in]}
\def\ITP{\footnote{Work supported by National Science Foundation
  grant PHY 89-08495.}\\[.1in] {\it Institute for Theoretical Physics\\
  State University of New York, Stony Brook, NY 11794-3840}\\[.1in]}
\def\endtitle{\par\end{quotation}\vskip3.5in minus2.3in\newpage}

\def\camera#1#2{
       \topmargin=.46in
       \textheight=22cm
       \textwidth=15cm
       \hsize=15cm
       \oddsidemargin=.28in
       \evensidemargin=.28in
       \marginparsep=0in
       \parindent=1.15cm
       \pagestyle{empty}
       \def\rm{\sf}
       \begin{document}
       \begin{center}{\Large\bf #1}\\[.5in minus.2in]{\bf #2}
       \vskip1in minus.8in {ABSTRACT}\\[.1in]\end{center}
       \renewcommand{\baselinestretch}{1}\small\normalsize
       \begin{quotation}\par}
\def\endabstract{\par\end{quotation}
       \renewcommand{\baselinestretch}{1.2}\small\normalsize}

\def\xpar{\par}                                       % \par in loops
\def\header{
 }
\def\letterhead{
 \header
 \font\sflarge=helvetica at 14pt   % if you have helvetica
 \leftskip=2.8in\noindent\phantom m\\[-.54in]
       {\large\sflarge STATE UNIVERSITY OF NEW YORK}
       {\scriptsize\sf INSTITUTE FOR THEORETICAL PHYSICS\\[-.07in]
       STONY BROOK, NY 11794-3840\\[-.07in]
       Tel: (516) 632-7979}
 \vskip.3in\leftskip=0in}
\def\letterneck#1#2{\par{\hbox to\hsize{\hfil {#1}\hskip 30pt}}\par
       \begin{flushleft}{#2}\end{flushleft}}
\def\letterhat{\parskip=\bigskipamount \def\baselinestretch{1}}% textile
\def\head#1#2{\letterhat\begin{document}\letterhead\letterneck{#1}{#2}}
\def\multihead#1#2{\thispagestyle{empty}\setcounter{page}{1}
       \letterhead\letterneck{#1}{#2}}        % multiple letters
\def\shead#1#2{\letterhat\begin{document}\sletterhead
       \letterneck{#1}{#2}}
\def\multishead#1#2{\thispagestyle{empty}\setcounter{page}{1}
       \sletterhead\letterneck{#1}{#2}}
\def\multisig#1{\goodbreak\bigskip{\hbox to\hsize{\hfil Kind regards,
       \hskip 30pt}}\nobreak\vskip .5in\begin{quote}\raggedleft{#1}
       \end{quote}}
\def\sig#1{\multisig{#1}\end{document}}

\def\watch{
 \newcount\hrs
 \newcount\mins
 \newcount\merid
 \newcount\hrmins
 \newcount\hrmerid
 \hrs=\time
 \mins=\time
 \divide\hrs by 60
 \merid=\hrs
 \hrmins=\hrs
 \divide\merid by 12
 \hrmerid=\merid
 \multiply\hrmerid by 12
 \advance\hrs by -\hrmerid
 \ifnum\hrs=0\hrs=12\fi
 \multiply\hrmins by 60
 \advance\mins by -\hrmins
 \number\hrs:\ifnum\mins<10 {0}\fi\number\mins\space\ifnum\merid=0
 AM\else PM\fi}
\def\half{\frac{1}{2}}

\def\eq{\begin{equation}}
\def\eqe{\end{equation}}
\def\eqa{\begin{eqnarray}}
\def\eqae{\end{eqnarray}}

\def\be{\begin{equation}}
\def\ee{\end{equation}}
\def\bea{\begin{eqnarray}}
\def\ena{\end{eqnarray}}

\def\to{\rightarrow}

\begin{document}
\def\1ov2{{1\over 2}}
\font\biggbold=cmbx10 scaled\magstep2
\font\bigbold=cmbx10 at 12.5pt
\font\bigreg=cmr10 at 12pt
%\nopagenumbers
\bigreg
\baselineskip = 18pt
\begin{flushright}
ITP-SB-95-49 \\
QMW-PH-95-40
\end{flushright}
\vskip 2.0cm

{\biggbold \centerline{The equivalence between the operator}}
{\biggbold \centerline{approach and the path integral approach for}}
{\biggbold \centerline{quantum mechanical non-linear sigma
models\footnote{
to appear in proceedings of workshop on gauge theories, applied supersymmetry
 and quantum gravity, Leuven, Belgium, July 10-14, 1995, and strings'95,
 USC, March 13-18, 1995.
}}}

\vskip 1.0cm

{\sc \centerline{Jan de Boer, Bas Peeters, Kostas Skenderis, and
Peter van Nieuwenhuizen}}

\baselineskip = 15pt
\centerline{Institute for Theoretical Physics,}
\centerline{State University of New York at Stony Brook}
\centerline{Stony Brook, NY 11794-3840, U.S.A.}
\vskip 1.2cm
\baselineskip=12pt

\noindent {\bf Abstract}:  We give background material and some
details of calculations for two recent papers [1,2] where we derived a
path integral representation of the transition element for supersymmetric
and nonsupersymmetric nonlinear sigma models in one dimension
(quantum mechanics).
Our approach starts from a Hamiltonian
$H(\hat{x}, \hat{p}, \hat{\psi}, \hat{\psi}^\dagger)$ with a priori
operator ordering. By inserting a finite number of complete sets of
$x$ eigenstates, $p$ eigenstates and fermionic coherent states,
we obtain the discretized path integral and the discretized propagators
and vertices in closed form.
Taking the continuum limit we read off the Feynman rules and measure of
the continuum theory which differ from those often assumed.
In particular, mode regularization of the continuum theory is shown
in an example to give incorrect results.
As a consequence of time-slicing,  the action and Feynman rules,
although without any ambiguities, are necessarily noncovariant, but the
final results are covariant if $\hat{H}$ is covariant.  All our derivations are
exact. Two loop calculations confirm our results.

\baselineskip=18pt
\section{Introduction.}
The subject of path integrals in curved space is arcane, complicated
and controversial [3].  In two recent articles we have considered
one-dimensional (quantum mechanical) path integrals, and found an
exact path integral representation for the transition element [1,2].  For the
bosonic case it is defined by
$T (z,y;\b)  = $ \\ $< z | \exp - {\b \over \hbar} \hat{H} ( \hat{x},
\hat{p}) | y >$ where $| y >$ and $< z|$ are position eigenstates.  The
classical Lagrangian is given by $L_{cl}= {1\over 2} g_{ij} (x) \dot{x}^i
\dot{x}^j$, but  both the quantum Hamiltonian and the
action in the path integral deviate substantially from $L_{cl}$.  We begin
by assuming that $\hat{H}$ has a  given a priori operator ordering.
By inserting $N-1$ complete sets of $x$ eigenstates and $N$ sets of $p$
eigenstates, one finds a discretized phase space path integral,  from
which one can derive (as we shall indeed do) discretized propagators
and vertices in closed form (by coupling to discretized external
sources).  In the continuum limit one obtains then a Euclidean path integral
of the form
 \eq
\int dp \; dx \; e^{ \int^0_{-\b} (i  p \dot{q} - H (p, q) \}  d t}
\label{intro1}
 \eqe
with well-defined propagators, vertices, and {\em rules how to
evaluate Feynman graphs}.  The last result is the most important:  these
rules are new and differ from what is usually assumed.

If one would bypass a detailed analysis of the discretized case, one
might expect that $L= i p \dot{q} - H (p, q)$ is covariant if $\hat{H}$ is
a covariant operator.  (For example, if $\hat{H}$ would commute with
the supersymmetry generators, one might expect that after
integrating out $p$, the resulting actions are the supersymmetric
actions one encounters in the literature).  This is incorrect:  the action
needs noncovariant terms of order $\hbar$ and $\hbar^2$ (but not
beyond) in order that $T$ be covariant.

Another source of puzzlement might be the observation that actions
of the form $L = {1\over 2} g_{ij} (x) \dot{x}^i \dot{x}^j$  contain
double-derivative interactions, leading to linearly divergent graphs by
power counting.  On the other hand, it is well-known that quantum
mechanics is a finite theory. It would seem strange (and is, in fact,
incorrect) to require that normal-ordering removes divergences:
where would normal ordering come from?  The resolution of this
paradox will be the presence of new ghosts, closely related to the
factors $g^{1/2} \d (0)$ which Lee and Yang found in a careful
treatment of the deformed harmonic oscillator [4], and which we shall call
for this reason ``Lee-Yang ghosts".  (They were first introduced by
Bastianelli, and in a more covariant form by him and one of us [5]).

The propagator for the quantum deviations of a scalar $q (\t)$  (where
$\t =  t/\b$) on the interval $(-1, 0)$ with boundary conditions $q (-1) = q
(0)=0$ is proportional to
\eq
\D (\s,\t) = \D_F (\s-\t) + (\s \t + {1\over 2} \s + {1\over 2} \t)
\label{intro1a}
 \eqe
where $\D_F (\s-\t) = {1\over 2} (\s-\t) \q (\s-\t) + \half (\t-\s) \q
(\t-\s)$ is the translationally invariant solution of $\del_\s^2 \D_F
(\s-\t) =
\d (\s-\t) $ while the terms $\s \t + {1\over 2} \s + {1\over 2} \t$
enforce the boundary conditions.  In Feynman graphs, contractions
between $x$ and $\dot{x}$, and between $\dot{x}$ and $\dot{x}$,
produce $\theta(\s-\t)$ and $\d (\s-\t)$,
and the problem arises how to evaluate products
of several $\d (\s-\t)$ and $\q (\s-\t)$. Mathematically, various
consistent but different definitions of these products of distributions can
be given [6], but physically (1) should have an unambiguous meaning, and
the problem is to find the correct rules.  Consider as an example
 \eq
\int^0_{-1} \; \int^0_{-1} \d (\s - \t) \q (\s - \t) \q  (\s - \t) d \s d \t
\label{intro1aa}
 \eqe
One might expect that, since $\d (\s - \t) = \del_\s  \q (\s - \t)$, the
result is equal to \\ ${1\over 3} \int^0_{-1} \; \int^0_{-1}\del_\s [\q
(\s-\t)^3] d \s dt = {1\over 3}$.  However, the correct result is ${1\over
4}$ as the discretized approach shows.

 Another source of ambiguities are the equal-time contractions, in
particular $<
\dot{x} (\t)
\dot{x} (\t)>$.  Are these the limit of
$< \dot{x} (\t_1) \dot{x} (\t_2) >$ for $\t_1 \to \t_2$,
and what to do with  the resulting $\d (0)$?  In field theory,
equal-time contractions are a priori undefined, and one needs a
symmetry principle to fix them, but here everything has been specified
from the beginning, so ambiguities in equal-time contractions are not
allowed.

Yet another worry would be the perennial headache called ``the
measure".  Since the path integrals correspond to a one-dimensional
quantum field theory on a finite time segment, one might expect some
factors in the measure to be present, like $\det g$ to some power at the
end points.  In fact, in Hamiltonian quantization of
gravity in higher dimensions, such factors (and factors involving
$g_{00}$) are present.  We shall see that also in our case there are
nontrivial measure factors.  These are usually omitted in calculations
with nonlinear sigma models, but are crucial to obtain the correct
results.

What do we exactly mean by ``correct results"?  ``Correct results" means
for us:  the results which agree with
$< z|\exp - {\b\over \hbar} \hat{H} | y>$,
no more and no less.  This matrix element can be straightforwardly
evaluated order by order in $\b$, without encountering any divergences or
ambiguities, simply by expanding the exponent and inserting a complete
set of $p$-states
 \eq
T (z, y; \b) = \int < z | \exp - {\b \over \hbar} \hat{H} \mid p >
< p | y > d^4p
\label{intro2}
 \eqe
Moving all $\hat{x}^j$ to the left and all $\hat{p}_j$ to the right,
keeping track of commutators, one obtains c-number results.  All
terms with a given number of commutators, say $s$, are of a given order in
$\b$ (see below), and although for each $s$ an infinite number of terms
contributes, one can sum the infinite series for  fixed $s$ in closed
form.  Thus $T(z, y; \b)$ is a finite and unambiguous Laurent series in
$\b$.  Our task is to find a path integral which reproduces these terms
order by order in loops (in the path integral, $\b$ and $\hbar$ appear
only in the combination $\b \hbar$, so $\b$ also counts the number of
loops on the worldline).  One could, of course, reject this Hamiltonian
starting point, and try to devise a more covariant way of defining path
integrals in curved space which does not require a given Hamiltonian
$\hat{H} (\hat{x}, \hat{p})$.  In particular, whereas in our approach we
encounter noncovariant ``midpoint rules" like
$\bar{x}_{k+1/2}={1\over 2} (x_{k+1} + x_k)$, one might hope that
covariant midpoint rules (the middle of a geodesic, for example) might
lead to a completely covariant treatment.  All we can say is that we
have found a path integral which yields the correct results and which
straightforwardly follows from $T (z, y; \b)$ by inserting complete
sets of states, whereas the more covariant approaches have so far
not been able to reach the same results\footnote{Covariant techniques
to evaluate path integrals do exist, but can only be used when the
path integral has additional special properties, for example if the
semi-classical approximation is exact, or if the theory has additional
symmetries so that one can use localization techniques. However,
such techniques do not extend to arbitrary path integrals. In order
to set up perturbation theory, one has to choose a decomposition
$S=S^{(0)}+S^{{\rm int}}$, and in the case of a general $\sigma$ model
it is not true that $S^{(0)}$ and $S^{{\rm int}}$ can be chosen
to be separately
covariant. (The kinetic operator in the full quadratic term is
covariant, but not invertible in closed form[3].) Thus, Feynman rules
based purely on $S^{(0)}$ are necessarily noncovariant.}.

There are good practical reasons for starting with the Hamiltonian
matrix element $< z | \exp - \b/\hbar H | y> $, rather than a covariant
configuration space starting point.  When one calculates anomalies in
$n$-dimensional quantum field theories, one can rewrite these
anomalies as products of matrix elements of the Jacobian times $T$[7],
\eq
An = \int \sqrt{g(y)} <y | J | z > \sqrt{g(z)} < z| \exp - {\b\over \hbar}
H | y > d^n y d^n z
\label{intro3}
 \eqe
The extension to include fermions is straightforward.  So, quantum
mechanics enters via $T$, and the reason one wants to rewrite $T$ as
a path integral is that the calculations of anomalies are much simpler
in the path integral approach than in the Hamiltonian approach.  We
shall not discuss these applications to anomalies here, but refer to [2].

We shall also not discuss fermions in detail here (again, see [2]), but
only say that one can introduce bras and kets $| \h >$ and $< \bar{\h} |$
which are eigenstates of the fermionic annihilation and absorption
operators $\hat{\j}^a$ and $\hat{\j}^\dagger_a$, respectively.  These
states are coherent states  $| \h > = (\exp \j^\dagger_a \h^a)|0>$
and  $< \bar{\h}| = < 0 |  \exp \bar{\h}_a \j^a$ with $\h^a$ and
$\bar{\h}_a$ {\it independent} Grassmann variables (so not related by
complex conjugation, only satisfying $\int d \h \h = \int d \bar{\h}
\bar{\h} = 1$).  This is the fermionic equivalent of the
``holomorphic" representation for bosonic systems [8].  We have found it
simplest to use the $x, p$ representation for coordinates, but for the
fermionic part the $\j ,\j^\dagger$ representation is by far the
most natural.   (One could, however, also construct a kind of $x, p$
representation for the fermions).  The fermionic coherent states
satisfy completeness relations and for $N=2$ supersymmetric systems
there is really no major obstacle to construct $T ( \bar{\h}, z , \h, y;
\b)$:  one combines $\j^a_\a (\a=1,2)$ into $\j^a \equiv (\j^a_1 + i
\j^a_2)/\sqrt{2}$ and $\j^\dagger_a = (\j^a_1 - i \j^a )/\sqrt{2}$.
However, for $N=1$ supersymmetric systems, one has $n$ Majorana
fermions
$\hat{\j}^a (a=1, \ldots , n)$ satisfying the Dirac brackets $\{ \hat{\j}^a,
\hat{\j}^b \} = \d^{ab}$.  To construct a vacuum and coherent states,
we need rather $\j^A$ and $\j^\dagger_A$ satisfying $\j^A | 0 > = 0, \{
\j^A, \j^B \} = 0 ,\{ \j^\dagger_A, \j^\dagger_B\} = 0$ and $\{ \j^A,
\j^\dagger_B \} = \d^A{}_B$.
This can be achieved in two ways: \linebreak
(i) by fermion doubling, namely adding a second set of free
fermions $\j^a_{II}$ which do not appear in the Hamiltonian but which
are used to construct $\j^A$ and $\j^\dagger_A$ as $\j^A=(\j^a_I + i
\j^a_{II})/\sqrt{2}$ and $\j^\dagger_A = (\j^a_I - i \j^a_{II}) /
\sqrt{2}$.  Here $\j^a_I$ denotes the original set of fermions $\j^a$.
\linebreak
(ii) by fermion halving, namely as follows
\eq
\j^A = (\j^{2A-1} + i
\j^{2A})/ \sqrt{2}, \j^\dagger_A = (\j^{2A-1} - i \j^{2A})/\sqrt{2}
\label{intro4}
 \eqe
The vacuum and Hilbert space are different in both cases, and one finds
different results for the propagators and transition elements (!), but
the anomalies come out the same. This is as expected:  in traces over
Hilbert space, differences created by choosing different vacua should
cancel.

\begin{figure}
\centerline{\hbox{\psfig{figure=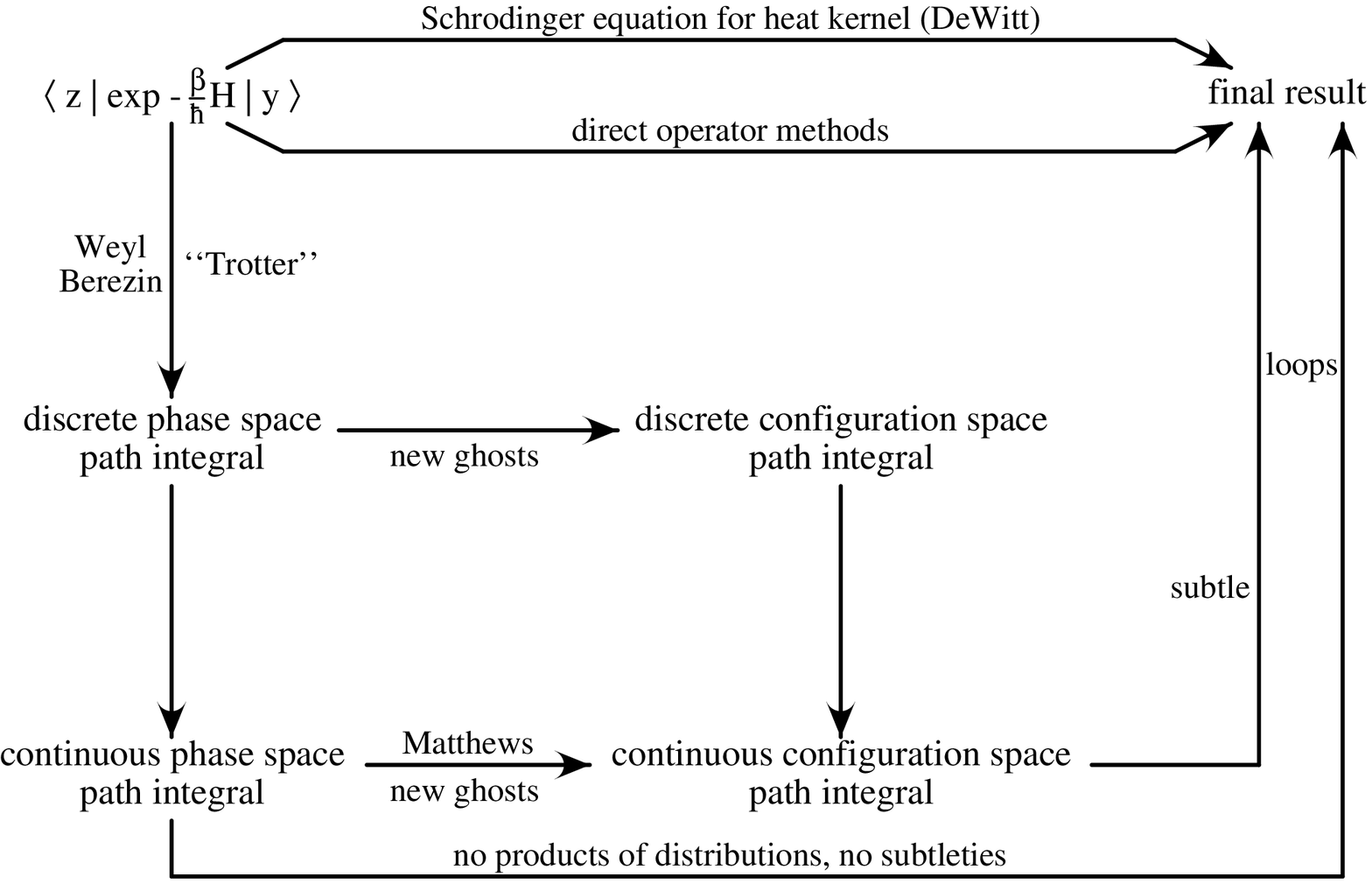,width=5in}}}
\end{figure}

In the next sections we shall give some details of calculations which
are only briefly summarized in [1,2].  A flow chart of the main ideas
is given above.
\nopagebreak

\section{Weyl ordering and an extension of the Trotter formula.}
For definiteness one may consider a particular Hamiltonian $\hat{H}_E$
which plays a role in the calculation of anomalies
 \eq
\hat{H}_E = {1\over 2} g^{-1/4} p_i g^{1/2} g^{ij} p_j g^{-1/4}
\label{five}
 \eqe
This operator is Einstein invariant, but we stress that our results hold for
any other operator with two $p$'s.  Inserting $(N-1)$ complete sets of  $x$
eigenstates and $N$ complete sets of $p$ eigenstates, using the
completeness relations $\int |x> \sqrt{g (x)} <~x| d^n x = \int |p><p|
d^n p = I$, we find an expression for $T (z, y; \b)$ in terms of $N$
kernels
 \eq
T (x_k, x_{k-1} ; \e) = \int  < x_k | \exp - {\e \over \hbar} \hat{H} | p_k
> < p_k | x_{k-1} > d^n p_k,
\label{six}
 \eqe
where $x_N = z, x_0=y$ and $\e = \b/N$.  We rewrite $\hat{H}$ as a Weyl
ordered operator.  For a polynomial in $p$'s and $x$'s, the
corresponding Weyl ordered operator is obtained by expanding $(\a^j
\hat{p}_j + \b_i \hat{x}^i)^N$ and retaining all terms with a particular
combination of $\a$'s and $\b$'s.  It follows that $(g^{ij} p_i p_j)_W=
{1\over 4} \hat{p}_i \hat{p}_j \hat{g}^{ij} + {1\over 2} \hat{p}_i
\hat{g}^{ij} \hat{p}_j + {1\over 4} \hat{g}^{ij} \hat{p}_i \hat{p}_j$, and
by evaluating
$\hat{H} - ({1\over 2} g^{ij} p_i p_j)_W$ one finds extra terms of order
$\hbar^2$
 \eq
\hat{H} = {1\over 2} (g^{ij} p_i p_j)_W + {\hbar^2 \over 8} (\G^\ell_{ik}
\G^k_{j\ell} g^{ij} + R)
\label{seven}
 \eqe
(on a sphere $R < 0 $).  Other operators $\hat{H}$ in which the $p$'s
appear less symmetrically will in general also lead to extra terms of
order $\hbar$.  For a polynomial one may prove that $(x^m
p^r)_W={1\over 2^m} \sum^m_{\ell=0} \left( \begin{array}{c} m \\
\ell \end{array} \right) \hat{x}^{m-\ell} \hat{p}^r \; \hat{x}^\ell$, and it
follows that
 \eqa
< z | (x^m p^r )_W | y > &=& \sum^m_{\ell = 0} {1\over 2^m} \left(
\begin{array}{c} m \\ \ell \end{array} \right) \int z^{m-\ell} p^r y^\ell < z
| p > < p | y > d^n p
\nonumber\\ &=& \int < z | p > \left( {z + y \over 2} \right)^m p^r < p | y >
d^n p
\label{eight}
 \eqae
This shows why orderings like Weyl ordering are very
convenient: one can replace
a Weyl ordered operator by a function, simply by substituting $\hat{p}
\to p, \hat{x} \to {1\over 2} (z+y)$,  {\em and this is an exact
result.}

However, Weyl ordering and exponentiation do not commute, $(\exp -
{\e \over \hbar} H)_W \not= \exp - {\e \over \hbar} (H_W)$, and
whereas $H_W$ was easy to write down, a closed expression for
$(\exp - {\e \over \hbar} H)_W$ cannot be written down.  One expects,
however, that a suitable approximation of the kernels, containing only
terms of order $\e$ suffices.  Here one stumbles upon a problem:
it might seem that $p$ is of order $\e^{-1/2}$ due to
the term $\exp - {1\over 2} \e p^2$ in the action.
Expansion of $\exp - {\e \over \hbar} H_W$
would contain terms of the form $\e^s p^r f(x)$ for which $s \geq 2$ but
which would still be of order $\e$.  We are now
going to give an argument that $p$ is of order unity, and therefore
only the terms with one explicit $\e$ need be retained.
Hence, we will use as kernel
$\exp - {\e \over \hbar} H_W ({1\over 2} (x_k + x_{k-1}),p_k)$.
In other words, the Trotter-like approximation
 \eq
< x | \exp - {\e \over \hbar} \hat{H} |p > \simeq  < x | 1 -  {\e \over \hbar}
\hat{H} | p > = (1-{\e\over \hbar} h) < x | p > = < x | p > \exp - {\e\over
\hbar} h
\label{nine}
 \eqe
is still correct, but $h$ is not simply $< x | \hat{H} | p>$ as in the usual
models with $H = T (p) + V (x)$,  but rather it equals $H_W$ at the
midpoints.  To obtain this result, we note that the kernels are exactly
equal to
 \eq
\int d^n  p_k \; e^{{i\over\hbar}  \vec{p}_k \cdot \vec{\D}x_{k-1/2}}
\bigg( e^{-{\e\over \hbar} H}\bigg)_W (\bar{x}_{k-1/2}, p_k) ;
\begin{array}{l} \D x_{k-1/2 } = x_k -x_{k-1} \\ \bar{x}_{k-1/2} = {1\over 2}
\left(  x_k + x_{k-1}\right) \end{array}
\label{ten}
 \eqe
The difference between $(\exp - {\e \over \hbar} H)_W$ and $\exp -
{\e \over \hbar} (H_W)$ consists of two kinds of terms
\begin{itemize}
\item [(i)] terms without a $p$ ; these are certainly of higher order in
$\e$ and can be omitted
\item [(ii)] terms with at least one $p$.
\end{itemize}
The crucial observation [9]
is now that the phase space propagators
$<p_{k, i} p_{l, j}>$ and $<p_{k, i} \bar{x}^j_{k+1/2}>$ are both
of order unity, and not of order $\e^{-1}$ and $\e^{-1/2}$, respectively.
A formal proof is given in equation (\ref{thirty2}). However,
already at this point one might note that the $p p$ propagator is not only
determined by the $g p p$ term but also by $i p \D q$. Completing squares,
it is the $p' = (p - i \D q/ \e)$ which is of order $\e^{-1/2}$.
In the $p p$ propagator the singularities of the $p' p'$ and
$\D q \D q$ propagators cancel each other. As a consequence, the
$p p$ and $p \bar{q}$ propagators are of order one, and this proves the
Trotter formula also for nonlinear sigma models.

The final result is that one may  use $\exp - {\e \over \hbar} H_W
({1\over 2} (x_k + x_{k-1}), p_k)$ as the kernels of the path integral.  If
one would not have used Weyl ordering, but simply computed $\int <
x_k  | \exp - {\e \over \hbar} \hat{H} | p_k> < p_k | x_{k-1}> d^n p_k$
keeping all terms of order $\e$, one finds terms proportional to $R_{ij}
(\bar{x}_{k-1/2}) \D x_{k-1/2}^i \D x^j_{k-1/2}$ where $R_{ij}$ is the Ricci
tensor.  These terms do not correspond to a local action (they are of the
form $\int R_{ij}
\dot{x}^i \dot{x}^j dt \e$).  These nonlocal kernels will yield the correct
answer for the path integral, but Feynman rules for nonlocal theories are a
headache, and it is not clear whether a truncation of these kernels to a
local action exists  which yields the correct answer.   Weyl ordering, on
the other hand, does lead to local kernels which are very easy to
construct and which yield the correct result.  This demonstrates the
usefulness of Weyl ordering.

\section{Discretized propagators and new Feynman rules.}
If we keep the $p_{k, j}$ and  $x^j_k$ as integration variables we obtain a
discretized phase space path integral, but if we integrate the $p_{k, j}$
out, we get a discretized configuration space path integral with $N$
factors
$[\det g_{ij} (\bar{x}_{k+1/2})]^{1/2}$.  In the continuum limit Lee and
Yang wrote these determinants as $\exp {1\over 2} \d (0) tr
\; ln \; g_{ij} dt$ and treated the exponent as a new term in the action [4].
We first discuss these configuration space path integrals.

For the same calculational advantages as in the case of Faddeev-Popov
ghosts in gauge theories, we exponentiate these determinants by
ghosts, but whereas the Faddeev-Popov determinant needs a pair of
anticommuting real ghosts/antighosts, here the square root of the
determinant requires an extra third commuting real ghosts.  One
obtains then the following path integral [5]
 \eqa
&\ &\prod^{N-1}_{k=0} (\det g_{ij} (\bar{x}_{k+1/2}))^{1/2} = \nonumber \\
&\ &\a \int db_{k+1/2}
dc_{k+1/2} da_{k+1/2}\exp {1\over 2} g_{ij} (\bar{x}_{k+1/2} ) \left[
b^i_{k+1/2} c^j_{k+1/2} + a^i_{k+1/2} a^j_{k+1/2} \right]
\label{eleven}
 \eqae
where $\a$ is a constant which can easily be determined.
We decompose the $x_k$ into a sum of background parts and quantum
parts, $x_k^j = x_{bg, k}^j  + q^j_k$, and the action $S$ into a free part
 \eq
S^{(0)} = \sum^{N-1}_{k=0} {1\over 2} g_{ij} (z) \left( \D q^i_{k+1/2}
\D q^j_{k+1/2} +b_{k+1/2}^i c^j_{k+1/2} + a^i_{k+1/2} a^j_{k+1/2} \right)
\label{twelve}
 \eqe
and an interaction part $S^{int}$ (the rest), requiring that the background
fields satisfy the (discretized) equation of motion of $S^{(0)} $ and the
boundary conditions.  Hence  $q_N = q_0 = 0.$ With the continuum limit in
mind we parametrize the discrete $q_k$ using continuum modes of
$S^{(0)}$
\be
q^j_k = \sum^{N-1}_{m=1} r^j{}_m \sqrt{2\over N} \sin {k m \p\over
N} ; k=1, \ldots , N-1 \label{mode}
\ee
The Jacobian for $x_k \to q_k \to r_m$ is unity.  We then couple  $\D
q_{k+1/2}^j$ and $\bar{q}_{k+1/2}^j$ to discretized external sources
 \eq
 S  ({\rm sources}) =- \sum^{N-1}_{k=0} (F_{k+1/2,j} \D q^j_{k+1/2} + \e
G_{k+1/2,} \; \bar{q}^j_{k+ 1/2}).
\label{thirteen}
 \eqe
Similarly we introduce discretized sources for the ghosts $b, c$ and $a$.
We then complete squares and integrate over the discrete variables
$r_m{}^j, b^j{}_{k + 1/2}, c^j{}_{k + 1/2}, a^j{}_{k + 1/2}$.  The result is
a functional quadratic in external sources which will yield the discretized
propagators.  We first quote the result and then give details of the
calculation.  By differentiating twice w.r.t $G$ one finds
 \eq
< \bar{q}_{k + 1/2}^i \bar{q}_{\ell + 1/2}^j > =
\e \hbar g^{ij} (z) \left[ -
(k+ {1\over 2} ) (\ell + {1\over 2}) / N +  (\ell + {1\over 2} ) \q_{k, \ell} +
(k + 1/2 ) \q_{\ell, k} \right]
\label{fourteen}
 \eqe
where $\q_{k, \ell}$ is the discretized $\q$ function
($\q_{k, \ell} = 0$ if
$k < \ell, \q_{k, \ell} = {1\over 2}\; {\rm if} \; k = \ell \; {\rm and} \;
\q_{k, \ell} = 1$ if $k>\ell$).  In the continuum limit this becomes
 \eqa
<  q^i   (\s) q^j (\t) > &=& - \b \hbar g^{ij} (z) \D (\s, \t) \; ; \;  -1
< \s = {k +
{1\over 2} \over N} - 1 < 0 \nonumber\\
\D (\s,  \t ) &=& \s (\t + 1 ) \q (\s - \t) + \t (\s + 1) \q (\t - \s)
\label{fifteen}
 \eqae
Similarly
 \eqa
<\bar{q}^i{}_{k+ 1/2}  \D q^j{}_{\ell + 1/2} > &=& \e \hbar g^{ij} (z)
\left[ - {k + \1ov2 \over N} + \q_{k, \ell} \right] \nonumber\\
< \D q^i{}_{k+1/2} \D q^j{}_{\ell + 1/2}>&=& {\e \hbar \over N} g^{ij}
(z) \left[ - 1 + N \d_{k, \ell} \right] \nonumber\\
< b^i{}_{k+ 1/2} c^j{}_{\ell+ 1/2} > &=& - {2\over \e}
\hbar g^{ij}(z) \d_{k, \ell} ; <
a^i{}_{k + 1/2} a^j{}_{\ell + 1/2} > = {1\over \e} \hbar g^{ij}(z) \d_{k, \ell}
\label{sixteen}
 \eqae

These results show that in the continuum limit $\q (\s-\t)= 1/2$ at $\s
= \t, \d (\s - \t)$ is a Kronecker delta even in the continuum theory
and not a Dirac delta, and they define equal-time contractions.  For
example
 \eq
< \dot{q}^i (\s ) \dot{q}^j (\s ) > + < b^i (\s ) c^j (\s) > + < a^i (\s)
a^j(\s) > = - \b \hbar g^{ij} (z)
\label{seventeen}
 \eqe
We see how the ghosts remove divergences, but we also see that a
well-defined finite part is left which in a less rigorous approach might
have been missed.  Terms in Feynman graphs with more than one $\d (\s-\t)$
are eliminated by the Lee-Yang ghosts whereas products of one $\d
(\s-\t)$ and any number of $\q (\s-\t)$ are evaluated by  still defining $\d
(\s - \t)$ in the continuum case to be a Kronecker delta.

\section{Derivation of the discretized propagators.}
The orthonormality of the matrix $O_k{}^m = \left({2\over
N}\right)^{1/2} \sin {1\over N} k m \p$  in (\ref{mode}) follows from
the formula $2
\sin
\a
\sin
\b =
\cos (\a - \b) - \cos   (\a + \b)$, and for $-2N < p < 2N$
 \eq
\sum^{N-1}_{m=1} \cos {p m \p \over N} = {1\over 2} \left(
\sum^N_{m=-N+1} e^{ipm\p/N} - 1 - (-)^p \right) = N\d_{p,0} - \half -
\half (-)^p
\label{eighteen}
 \eqe
The free action $S^{(0)}$ is diagonal in $r$'s since the $O$'s in $\D
q_{k-{1\over 2}} \D q_{k-{1\over 2}}$  in (\ref{twelve}) appear as
 \eq
\sum^N_{k=1} \left( O_k{}^m - O_{k-1}{}^m \right) \left( O_k{}^n -
O_{k-1}{}^n \right) = 2 \d^{mn} - \sum^N_{k=1} O_k{}^m \left(
O_{k-1}{}^n + O_{k+1}{}^n \right)
\label{nineteen}
 \eqe
and using $\sin \a + \sin \b = 2 \sin \half (\a + \b) \cos \half (\a - \b)$,
the orthogonality of $O_k{}^m$ leads to
 \eq
S^{(0)} = {1\over \e} \sum^N_{m=1} g_{ij} (z) r^i_m r^j_m \left(1 - \cos
{m
\p \over N} \right)
\label{twenty}
 \eqe
Adding $S$ (sources) to $S^{(0)}$, we find
 \eqa
&& Z \left[ \{ F \}, \{ G \} \right] = \int \prod^n_{i=1}
\prod^{N-1}_{m=1} d r_m{}^{i} \exp - {1 \over \hbar}  \left[ S^{(0)} +
\right. \nonumber\\
&&\left. \sum^{N-1}_{k=1} \sum^n_{j=1} \left\{ {1\over \e} (F_{k-1/2, j} -
F_{k+ 1/2, j} ) + \half \left( G_{k-1/2, j}+ G_{k+1/2, j}\right) \right\}
\left\{ \sum^{N-1}_{m=1} \sqrt{{2\over N}} r_m{}^j \sin {k m \p \over N}
\right\} \right] \hphantom{xxxx}
\label{twenty1}
 \eqae
Completing squares and integrating over $r_m{}^j$ yields
 \eqa
Z &=& \left[ \prod^{N-1}_{m=1} {(\p \e \hbar)^{n/2} \over \det g
(z)^{1/2} (1-\cos {m\over N} \p)^{n/2}}\right] \exp \left[
\sum^{N-1}_{m=1} \frac{\e \hbar}{4 (1-\cos {m \p \over N})} \right]
\O (F, G)^2 \nonumber\\
\O_j (F, G) &=& {2\over \e} \sqrt{{2\over N}} \sin {m \p \over 2N}
\sum^{N-1}_{k=0} \cos (k + \half ) {m \p \over N} F_{k + 1/2, j}
\nonumber\\
&& + \sqrt{{2\over N}} \cos {m\p \over 2N} \sum^{N-1}_{k=0} \sin (k+\half )
{m\p \over N} G_{k+ 1/2, j}
\label{twenty2}
 \eqae
The square of $\O$ is, of course, taken with $g^{ij} (z).$

The easiest propagator to compute is $<\dot{q} \dot{q} >$.  By
differentiation w.r.t. $F_{k+1/2, i}$ and $F_{\ell + 1/2, j}$  one finds
that the square of $\sin {m\p \over 2N}$ cancels the factor $(1-\cos
{m\p \over N})$ in the denominator, and using $\cos \a \cos \b = \half
\cos (\a + \b) + \half \cos (\a - \b)$, one must evaluate the sums
$\sum^{N-1}_{m=1}$ of $\cos (k + \ell + 1)m\p/N$ and $\cos (k-\ell)
m\p/N$, for which one may use (\ref{eighteen}).  The result is given in
(\ref{sixteen}).

Next we consider the $< q \dot{q}>$ propagator.  Differentiation w.r.t.
$G_{k+1/2, i}$ and $F_{\ell + 1/2, j}$ leads to a product $\cos {m\p \over
2N} \cos (\ell + \half ) {m \p \over N} \sin (k +
\half ) {m \p \over N} \sin {m \p \over 2N}.$   The last factor partly cancels
the denominator $1-\cos {m\p \over N}$.  One must then evaluate the
series
 \eqa
&& \sum^{N-1}_{m=1} \cos {m\p \over 2N} \left({\sin (k + \half) {m\p
\over N}\over \sin {m\p \over 2N}} \right) \cos (\ell + \half ) {m \p
\over N} \nonumber\\
&& = {1\over 4} \sum^{N-1}_{m=1} ( \z^m + \z^{-m}) (\z^{2km} +
\z^{(2k-2)m} + \cdots +\z^{-2km} ) \left( \z^{(2\ell + 1)m} +
\z^{-(2\ell + 1)m} \right)
\label{twenty3}
 \eqae
where we defined $\z = \exp {i\p\over 2N}$.  We write this series as a sum
of four series, and combine terms pairwise such that we can use
(\ref{eighteen}).  The
first two series start with $\z^{(2k+2 \ell  + 2)m}$ and $\z^{(2k + 2 \ell)m}$,
respectively, and run till $\z^{(-2k+2\ell+2)m}$ and
$\z^{(-2k+2\ell)m}$, while the last two series we write in ascending
order such that they start with $\z^{(-2k -2\ell -2)m}$ and
$\z^{(-2k-2\ell)m}$ and run till  $\z^{(2 k - 2 \ell -2)m}$ and
$\z^{(2k-2\ell)m}$, respectively.  The terms  in the
first and third series are pairwise combined using (\ref{eighteen}), and
similarly the terms in the second and fourth
series.  One finds then
\eq
{1\over 4} \sum^{k+\ell +1}_{p=- k+\ell +1} \left( -1- (-)^p + 2 N \d_{p,
0} \right) + {1\over 4} \sum^{k+\ell}_{p=-k+\ell} \left( -1 - (-)^p + 2 N
\d_{p,o} \right)
\label{twenty4}
 \eqe
The terms with $(-)^p$ cancel.  In the remainder one may distinguish
the cases $k > \ell, k < \ell$ and $k=\ell$.  One finds then ${1\over 4}
\left[ -2(2 k+1) + 2N \d_{k \geq \ell} + 2 N \d_{k>\ell} \right]$.
So the result is proportional to $(k+1/2)/N + \half \d_{k \geq \ell} +
\half \d_{k > \ell}$ which agrees with (\ref{sixteen}).

Finally we consider the $qq$ propagator.  This is the most
complicated one.  Differentiation with $G_{k + 1/2, i}$ and $G_{\ell +
1/2, j}$ leads to the series
 \eqa
\sum^{N-1}_{m=1} (\cos {m\p \over 2N})^2 \; \left( {\sin (k+ 1/2) {m\p
\over N}\over \sin {m\p \over 2N}} \right) \; \left( {\sin (\ell + \half)
{m\p \over N} \over \sin {m\p \over 2N}} \right)
\label{twenty5}
 \eqae
Again we rewrite this as series in $\z$
 \eqa
&& \sum_{m=1}^{N-1} (\z^m + \z^{-m})^2 (\z^{2km} +
\z^{(2k-2)m} + \cdots + \z^{-(2k-2)m} + \z^{-2km}) (\z^{2\ell m} + \z^{(2\ell
-2)m}  + \cdots +
\z^{-2 \ell m} ) \nonumber\\
&=& \sum^{N-1}_{m=1} \sum^k_{\a = - k} \sum^\ell_{\b = -
\ell} \left(
\z^{(2 \a + 2 \b + 2)m} + 2 \z^{(2 \a + 2 \b)m} + \z^{(2 \a + 2 \b - 2)m}
\right)
\label{twenty6}
 \eqae We combine $\z^{(2 \a + 2 \b +2)m}$ in the first series with
$\z^{(-2 \a - 2 \b - 2)m}$ of the last series, and $\z^{(2\a + 2\b)m}$
with $\z^{(-2
\a - 2 \b)m}$.  Then (\ref{eighteen}) yields
 \eq
\sum^k_{\a = - k} \sum^\ell_{\b=-\ell} \left[ \left( - \half - \half
(-)^{\a + \b + 1} + N\d_{\a + \b + 1, 0} \right) + \left( - \half - \half
(-)^{\a+\b} + N
\d_{\a +
\b, 0}\right) \right]
\label{twenty7}
 \eqe The summand becomes $-1+N ( \d_{\a + \b + 1, 0} + \d_{\a + \b,
0})$ and considering separately the cases $k > \ell, k < \ell$ and
$k=\ell$ we find
 \eqa - (2k+1)(2\ell +1) + \begin{array}{l} 2N (2\ell + 1) \; {\rm for} \; k
> \ell
\\ N (4 k +1) \; {\rm for} \; k = \ell \\ 2 N ( 2 k + 1) \; {\rm for} \; k
< \ell
\end{array}
\label{twenty8}
 \eqae
This agrees with (\ref{fifteen}).

\section{Higher loop calculations.} The transition element is
now given by
 \eq
 T  (z, y ; \b) = \left({g (z)\over g (y)} \right)^{1/4} \; \left( \exp -
{1\over \hbar} S^{int} \right) (\exp - {1\over \hbar} S^{prop})
\label{twenty9}
 \eqe
The factor $\left\{ g (z) / g (y)\right\}^{1/4}$ comes from
our choice of free and interaction part and our
normalization of states, in particular,  $< y | p > = (2\p\hbar )^{-n/2}
g(y)^{-1/4} \exp {i\over \hbar} p_j y^j$.  This nontrivial measure factor is
usually omitted but is crucial to get correct results.
Vertices are given by
 \eqa
{1\over \hbar} S^{int} &=& {1\over \b \hbar} \int^0_{-1} \left[ \right.
\half g_{ij} (x_{bg} + q) \left\{ (\dot{x}^i_{bg} + \dot{q}^i)
(\dot{x}_{bg}^j + \dot{q}^j) + b^i c^j + a^i a^j \right\} \nonumber\\
&-& \half g_{ij} (z)
\left\{ \dot{q}^i \dot{q}^j + b^i c^j + a^i a^j
\right\}  \left. \right] d \t \nonumber\\ &-& {1\over 8} \b \hbar
\int^0_{-1} (\G \G + R) d \t  \;  ; \; x_{bg} (\t ) = z + (z-y) \t
\label{thirty}
 \eqae
and propagators are given in (15-16).  One can now
compute the loop expansion of $T$;  this involves higher loops of a
quantum field theory on a finite time segment.  Our lattice regularization
defines all expressions and the results are finite, unambiguous and
correct (see section 1 for the definition of correct). The two-loop
corrections to $T$ agree with the result obtained from direct operator
methods (see the flow chart).

\section{Phase space path integrals.}

If one moves in the flow chart down on the left hand side, one
encounters phase space path integrals.  Coupling the nN momenta
$p_{k, j}$ to external sources
$F^j{}_{k-{1\over 2}}$ ($p_k$ lies between $x_{k}$ and $x_{k-1}$, and will
become equal to $i \D q_{k - 1/2}/\e$.  We are in the Euclidean case), and
the midpoint fluctuations
$\bar{q}^j_{k-1/2}$ to $G_{k -1/2, j}$ one finds, after completing
squares and integrating out the $p$'s, in the exponent a factor
\eq
\sum^{N-1}_{k=0} {\e \over 2 \hbar} \{ - i F^j{}_{k+1/2} + {i\over \e}
(q^j{}_{k+1}  - q_k^j) \}^2
 \label{thirty1}
 \eqe Expanding this term, one recovers the result already obtained for
the discretized configureation space path integral, together with an
extra $F^2$ term.  It follows that the $\bar{q} \bar{q}$ and $\bar{q} p$
propagators in phase space are the same as the $\bar{q}\bar{q}$ and
$i \bar{q} \dot{q}$ propagators in configuration space, but the $pp$
propagator is equal to minus the
$\dot{q}\dot{q}$ propagator plus an extra term proportional to $\d_{k,
\ell}$, which cancels the $\d_{k, \ell}$ present in $< \dot{q}\dot{q}>$.
Hence, the $p$ propagator is nonsingular.  No $\d (\s - \t)$ are present in
continuum phase space Feynman graphs, and the naive approach gives the
correct results

 \eq < p_i (\s) p_j (\t) > = \b \hbar g_{ij} (z) ; < q^i (\s ) p_j (\t ) >
= - i \b \hbar \d_j^i (\s + \q (\t - \s) )
\label{thirty2}
 \eqe

The naive propagator for the kinetic terms  ${\b
\over \hbar} \int^0_{-1} [ i p_i \dot{q}^i  - \half g^{ij}(z) p_i p_j ] d\s$ is
given by $ G (\s, \t) = \left(\begin{array}{cc} 1 & - i \del_\s \\ + i \del_\s
&
0
\end{array}
\right)^{-1} \d(\s - \t)$.    We decompose $G$ as
$G (\s, \t) = G_F (\s - \t) + P (\s, \t) $  where $P$ is annihilated by the
field
operator (the homogeneous solution) and $ G_F (\s - \t)  = \left(
\begin{array}{cc} 0 & + \half i \e (\s - \t) \\ - \half i \e (\s-\t) &  \D_F
(\s-\t)
\end{array} \right)$.  The boundary conditions $q (\s = 0) = q (\s = - 1) = 0$
fix $P$ completely, and one recovers (\ref{thirty2}).  Note that one does
not need any boundary conditions on $p$, nor is there any need, since there
are no zero modes in $p$: all $p$ integrals are convergent and Gaussian.
Our discretized approach explains this:   the variables $p_k$ were defined
at midpoints (between $x_k$ and $x_{k-1}$, and were not specified at the
endpoints, unlike the $x_k$ for which $x_N = z$ and
$x_0=y$.  Similar remarks hold for the ghosts:  also they are defined on the
midpoints, have no boundary conditions and the $b,c$ integrations
converge because these are Grassmann integrations while the $a$
integrations converge because they are Gaussians.

\section{Mode regularization.}

In this section we will illustrate that the commonly used mode cut-off
regularization scheme gives incorrect results for the transition element.
In the mode-cut off regularization
scheme one starts directly from the continuum configuration path integral.
All quantum fields are expanded in a Fourier series and the path integral
is converted into an integral over the Fourier modes. The mode regularization
scheme amounts to performing all the calculations with a fixed number of
Fourier modes, say $M$, and then at the end of the calculation let
$M \rightarrow \infty$.  We shall see that this seemingly ``natural''
regularization scheme is inconsistent with out new Feynman rules
and therefore yields incorrect result (incorrect in the sense explained in the
introduction).

\begin{sloppypar}
To be concrete let us consider the same model as before.
The continuum action is given by
\eqa
S &=& {1\over \b} \int^0_{-1}
\half g_{ij} (x_{bg} + q) \left[ (\dot{x}^i_{bg} + \dot{q}^i)
(\dot{x}_{bg}^j + \dot{q}^j) + b^i c^j + a^i a^j \right]
\nonumber \\
&-& {1\over 8} \b \hbar
\int^0_{-1} (\G \G + R) d \t  \;  ; \; x_{bg} (\t ) = z + (z-y) \t
\eqae
The background fields $x_{\rm bg} (\t)$
satisfy the field equation of the quadratic part
$S^{(0)}= {1\over 2 \b} g_{ij} (z) \int^0_{-1}
(\dot{q}^i \dot{q}^j + b^i c^j + a^i a^j) d \t$
and are chosen such that they vanish at the boundary.
\end{sloppypar}

Since the quantum fields $q^i(\t)$ vanish at the boundary we can expand them in
the complete set of $\{ \sin n \p \t \}$ on the interval $-1 \leq \t \leq 0$.
The ghosts we expand into $\cos n \p \t$ since they don't vanish at the
boundaries
\bea
&\ & q^i = \sum_{n=1}^{\infty} q^i_n \sin (n \p \t); \ \
b^i = \sum_{n=0}^{\infty} b^i_n \cos (n \p \t) \nonumber \\
&\ & c^i = \sum_{n=0}^{\infty} c^i_n \cos (n \p \t); \ \
a^i = \sum_{n=0}^{\infty} a^i_n \cos (n \p \t).
\eqae
Next we change variables in the path integral from the quantum fields
to modes. At this stage the measure is fixed by hand such that a Gaussian
integral over each mode gives one (apart from a possible overall constant).
It is straightforward to obtain the propagators
\bea
<q^i(\s) q^j(\t)> &=& - \b \hbar g^{ij} \D(\s, \t), \label{xpro} \\
<b^i(\s) c^j(\t)> &=& - 2 \b \hbar g^{ij} \pa_\s^2 \D(\s, \t),\label{bpro} \\
<a^i(\s) a^j(\t)> &=&  \b \hbar g^{ij} \pa_\s^2 \D(\s, \t),\label{apro}
\eqae
where
\be
\D(\s, \t) = -2 \sum_{n=1}^\infty \frac{\sin(n \p \s) \sin(n \p \t)}{n^2 \p^2}
\label{delta}
\ee
Note that (\ref{bpro}) and (\ref{apro}) follow from the identity
$2 \sum_{n=1}^\infty \cos n \p \s \cos n \p \t + 1 =
2\sum_{n=1}^\infty \sin n \p \s \sin n \p \t = \d(\s -\t)$. (Use that
$\theta(\s - \t)
= - 2 \sum_{n=1}^\infty \cos n \p \s \sin n \p \t/(n \p) - \t$
is also given by
$2 \sum_{n=1}^\infty \sin n \p \s \cos n \p \t/(n \p) + \s +1$,
and differentiate w.r.t. $\s$). From this identity (\ref{seventeen}) follows.
In fact, expanding the ghosts into sines gives the same propagators, as the
identity shows.

The propagators
$<\dot{q}^i(\s) q^j(\t)>$ and $<q^i(\s) \dot{q}^j(\t)>$ and
$<\dot{q}^i(\s) \dot{q}^j(\t)>$ are obtained  by
simply differentiating (\ref{xpro}) appropriately.
Mode cut-off regularization means that we truncate $\D(\s, \t)$
at some mode $M$, perform all calculations and at the
end let $M \rightarrow \infty$.

Let us now illustrate that mode regularization yields incorrect results.
Consider the two loop graph with contribution
\be
J = \int \int d\t d\s \D^.(\s, \t) \ {}^.\D(\s, \t) \ {}^.\D^.(\s, \t),
\ee
where the dot in $\D(\s, \t)$ indicates a time derivative w.r.t. $\s$ or
$\t$ depending on which side the dot is (for example
$\D^.(\s, \t) = \pa_\t \D(\s, \t)$).

Using (\ref{delta}) and performing the integrals over $\s$ and $\t$ we get
\bea
&J& = -\frac{1}{\p^4} \sum_{m,n,k=1}^{\hspace{.7cm} \prime}
\frac{1-(-1)^{m+n+k}}{mn}
\left[ \left (\frac{1}{m+n+k} + \frac{1}{m+n-k} \right)^2 \right. \nonumber \\
&\ &\hspace{4cm}
- \left. \left (\frac{1}{m-n+k} + \frac{1}{m-n-k} \right)^2 \right],
\eqae
where the prime indicates that we only sum over $m, n, k$ such that
all denominators are nonzero. This triple sum is only conditionally
convergent. Its result depends on the way the summation is performed.
Mode cut-off instruct us that we perform all sums for a finite upper
limit $M$ (the same for all three) and then let the cut-off tend to infinity.
A numerical calculation yields $\frac{-1}{12}$, whereas our Feynman rules
give $\frac{-1}{6}$. Clearly mode cut-off is incorrect for this problem.

\section{Outlook.}

Our results for nonlinear sigma models can serve as
a toy model for higher-dimensional path-integrals, to clarify there such
problems as:  equal-time contractions, higher-derivative interactions,
the measure, boundary conditions, extra ghosts.  In particular the role
of the extra terms due to Weyl ordering is intriguing.  If one follows
Schwinger's analysis of Yang-Mills theory in the Coulomb gauges [10], one is
dealing with a four-dimensional nonlinear sigma model. The operator
ordering in the Hamiltonian may be fixed by starting with Yang-Mills
theory in the $A_0=0$ gauge (where no ordering ambiguities exist and
where it seems therefore reasonable to take the Hamiltonian operator
without extra $\hbar$ terms) and then to make a canonical
transformation (at the quantum level!) to the Coulomb gauge.  This
produces extra terms of order $\hbar$ and $\hbar^2$ in the Coulomb
Hamiltonian which Schwinger already discovered by requiring that the
Poincar\'e generators close at the quantum level.  According to
Christ and Lee [11], Weyl ordering will
lead to further $\hbar$ and $\hbar^2$ corrections.

On the other hand, the configuration space approach of Faddeev and
Popov also ends up with Feynman rules for the same theory, but here
there is no sign of
$\hbar$ corrections and the Feynman rules are straightforward.  In fact,
the FP approach is  only intended to yield the Feynman rules at the $\hbar =
0$ level, but it does not address itself to
$\hbar$ corrections.  Yet, the Coulomb gauge plays a central role in
fundamental (not practical) discussions of quantum gauge field theory
and a precise understanding of the quantum theory requires to settle
issues at order $\hbar$ and beyond. It would therefore be very interesting
to generalize our framework and to establish
a well-defined set of Feynman rules in higher dimensions.
The question then arises whether our
Hamiltonian approach is equivalent to the naive FP
method. If not, this might have profound implications.

\baselineskip=15pt
\noindent {\bf References.}
\begin{enumerate}
\item J. de Boer, B. Peeters, K. Skenderis and P. van Nieuwenhuizen, {\it
Nucl. Phys. B} {\bf 446} (1995) 211, hep-th/9504087.
\item J. de Boer, B. Peeters, K. Skenderis and P. van Nieuwenhuizen, to
appear in {\it  Nucl. Phys. B}, hep-th/9509158.
\item B. De Witt, ``Supermanifolds", 2nd edition, Cambridge University
Press, 1992; L. Schulman, ``Techniques and applications of path
integratiion", John Wiley and Sons, New York, 1981.  Chapter 24 gives a
review.
\item T.D. Lee and C.N. Yang, {\it Phys. Rev. D} {\bf 128} (1962), 885.
\item F. Bastianelli, {\it Nucl. Phys. B} {\bf 376} (1992) 113; F. Bastianelli
and P. van Nieuwenhuizen, {\it Nucl. Phys. B} {\bf 389} (1993) 53.
\item J.F. Colombeau, {\it Bull. A.M.S.} {\bf 23} (1990) 251, and J.F.
Colombeau, ``Multiplication of distributions", Lecture Notes in
Mathematics, 1532, Springer-Verlag.
\item L. Alvarez-Gaum\'e and E. Witten, {\it Nucl. Phys. B} {\bf 234} (1989)
269.
\item L. Faddeev and A. Slavnov, ``Gauge Fields:  an Introduction to
Quantum Theory", 2nd ed., Addison-Wesley, Redwood City, 1991.
\item K.M. Apfeldorf and C. Ordo\~{n}ez, `Field redefinition invariance
and ``extra'' terms', UTTG-29-93, hep-th/9408100.
\item J. Schwinger, {\it Phys. Rev.} {\bf 127} (1962) 324; 130 (1963) 406.
\item N.H. Christ and T.D. Lee, {\it Phys. Rev. D} {\bf 22} (1980) 939.
\end{enumerate}

\end{document}